\documentclass[aps,prc,twocolumn,tightenlines,showpacs,
superscriptaddress,nofootinbib,14pt]{revtex4}
\usepackage[figuresright]{rotating}

\begin{document}

\title{Enrichment of the Superheavy Element Rg in Natural Au}

\author{A. Marinov}
\email{marinov@vms.huji.ac.il} \altaffiliation{Fax:
+972-2-6586347.}
 \affiliation{Racah Institute of Physics, The Hebrew
University of Jerusalem, Jerusalem 91904, Israel}
\author{A. Pape}
\affiliation{IPHC-UMR7178, IN2P3-CNRS/ULP, BP 28, F-67037
Strasbourg cedex 2, France}
\author{D. Kolb}
\affiliation{Department of Physics, University GH Kassel, 34109
Kassel, Germany}
\author{L. Halicz}
\affiliation{Geological Survey of Israel, 30 Malkhei Israel St.,
Jerusalem 95501, Israel}
\author{I. Segal}
\affiliation{Geological Survey of Israel, 30 Malkhei Israel St.,
Jerusalem 95501, Israel}
\author{N. Tepliakov}
\affiliation{Geological Survey of Israel, 30 Malkhei Israel St.,
Jerusalem 95501, Israel}
\author{Y. Kashiv}
\affiliation{Racah Institute of Physics, The Hebrew University of
Jerusalem, Jerusalem 91904, Israel}
\author{R. Brandt}
\affiliation{Kernchemie, Philipps University, 35041 Marburg,
Germany}

\date{November 9, 2010}

\pacs{21.10.Dr, 21.10.Tg, 27.90.+b, 36.10.-k}

\begin{abstract}

Based on the observation of the long-lived isotopes $^{261}$Rg and $^{265}$Rg (Z = 111, t$_{1/2}$ $\geq$ 10$^{8}$ y) in natural Au, an experiment was performed to enrich Rg in 99.999\% Au. 16 mg of  Au were heated in vacuum for two weeks at a temperature of 1127$^\circ$ C (63$^\circ$ C above the melting point of Au). The content of $^{197}$Au and $^{261}$Rg in the residue  was studied with high resolution  inductively
coupled plasma-sector field mass spectrometry (ICP-SFMS). The residue of Au was 3x10$^{-6}$ of its original quantity. The recovery of Rg was a few percent. The abundance of Rg compared to Au in the enriched solution was  about 2x10$^{-6}$, which is a three to four orders of magnitude enrichment. It is concluded that the evaporation rate of Rg from an Au matrix in vacuum at 63$^{o}$ C above the Au melting point is lower than that of Au. This experiment reinforces our first observation of Rg in a terrestrial material. As before it is concluded that a long-lived isomeric state exists in $^{261}$Rg and that it probably belongs to a new class of isomeric states, namely high spin super- or hyperdeformed  isomeric states.
\end{abstract}

\maketitle
In recent years long-lived isomeric states have been reported in several isotopes found in natural material \cite{mar07,mar09,mar08}. They, for instance, were  seen  in
 the neutron deficient  $^{211}$Th, $^{213}$Th, $^{217}$Th
and $^{218}$Th isotopes, with estimated half-lives  of
$\geq$ 1x10$^{8}$ y, which is 16 to 22 orders of magnitude longer
than the known half-lives of their corresponding g.s. \cite{mar07,fire96}.
They were also observed in the superheavy element region \cite{mar09,mar08}.
Isotopes with atomic mass numbers 261 and 265 have been reported in natural Au \cite{mar09} and interpreted
as long-lived isomeric states in $^{261}$Rg and $^{265}$Rg, since Rg (Z = 111) is a
chemical homolog of Au \cite{sea68}. In the superactinides, a nucleus with atomic mass number  292 has been found in purified Th and interpreted as most probably a nucleus with atomic number 122, because element 122 is predicted to be eka-Th \cite{sea68,eli02}.  Since all these nuclei were observed in natural materials it was deduced that the lower limit on their half-lives should be about 1x10$^{8}$ y, or otherwise they would have decayed away. The predicted half-lives for $^{261}$Rg, $^{265}$Rg and $^{292}$122 in their normally deformed  g.s. are  in the region of 10$^{-6}$ to 10$^{-8}$ s \cite{mol97}. Therefore it was concluded that, like in the neutron-deficient Th isotopes, these nuclei are also in long-lived isomeric states. It was hypothesized that they may belong to a new class of  high spin superdeformed (SD) or hyperdeformed (HD)  isomeric states. Such isomeric states have been seen in $^{195}$Hg \cite{mar01a}, in $^{210}$Fr \cite{mar96a,mar96b} and in heavy actinide nuclei like $^{236}$Am, $^{236}$Bk \cite{mar87}, $^{247}$Es and $^{252}$No \cite{mar01b}. Their character as high spin SD and HD  isomeric states has been proven by observation of abnormal $\alpha$-particle groups in coincidence with $\gamma$-rays of   SD bands \cite{mar01a,mar96a}.

In the experiments of \cite{mar09,mar08} it was assumed that the chemical procedures used by the purveyors of separated  Au and Th were not so specific as to separate the heavy homologs Rg and eka-Th from Au and Th, respectively. The purpose of  the present experiment was to try to concentrate Rg in Au. If the evaporation rate of Rg  is lower than that of Au from a melt, it may be possible to enrich Rg in Au by keeping the Au in vacuum a little over its melting point. One may thus preferentially evaporate Au relative to Rg. (The melting point of Au is 1064$^\circ$ C and that of Ag, its lighter homolog, is 961$^\circ$ C. The corresponding boiling points, under normal pressure are 2866$^\circ$ C and 2162$^\circ$ C, respectively.)

As before \cite{mar07,mar09,mar08} accurate mass measurements, using ICP-SFMS, have been employed on mass 261 to identify $^{261}$Rg. The mass M$_{A}$ of an atom is equal to:\\

M$_{A}$ = ZM$_{H}$ + NM$_{n}$ - BE \hspace{3.5 cm} (1)\\

where Z and N are the number of protons and neutrons in the atomic nucleus, M$_{H}$ and M$_{n}$ are the masses of the hydrogen atom and of the neutron, respectively, and BE is the binding energy of the atom.    The binding energy per nucleon (BE/u) as a function of A for stable nuclei has a maximum at about A = 60 and then drops monotonically towards higher A \cite{eva55}. Therefore, the mass of a heavy atom is higher than and separated from the mass of any molecule (except for multi-H, -Li, -Be and -B molecules) with the same mass number.

            \begin{figure}[h]

\includegraphics[width=0.47\textwidth]{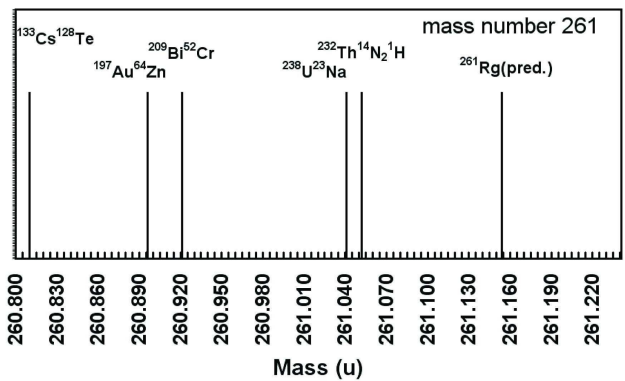}
 \caption{Representation of the systematic behavior of the masses
 of various M = 261 species, ranging from the quasi-symmetric
 combination $^{133}$Cs$^{128}$Te to Au- Bi-, Th-
 and U-based molecules \cite{aud03}, to
 the predicted \cite{mol95,lir00,kou05} mass of $^{261}$Rg.}
\end{figure}

This is seen in
Fig. 1, where the systematic behavior of the masses \cite{aud03}
of various M = 261 species, from the quasi-symmetric combination
$^{133}$Cs$^{128}$Te to Au-, Bi-, Th- and U-based molecules to the
predicted \cite{mol95,kou05,lir00} mass of the $^{261}$Rg  nucleus is
  displayed. The mass of the very
neutron-rich $^{261}$Au  nucleus is predicted to be
261.337 u \cite{mol95} or 261.324 u \cite{kou05}, values well above the expected
mass of the $^{261}$Rg nucleus, which is  261.154 u
 \cite{mol95,lir00,kou05}. (The calculations \cite{mol95,lir00,kou05} of the
$^{261}$Rg mass agree to within  0.002 u.) Thus, as was demonstrated before \cite{mar09,mar08},
accurate mass measurements are an effective tool for detecting
naturally-occurring superheavy elements.

In the present experiment, an Au solution enriched in Rg has been produced. Accurate mass measurements in the region of mass 261 were performed  using ICP-SFMS with a resolution  m/$\Delta$m = 4000. A peak that fits with the predicted mass of $^{261}$Rg and does not fit with any known molecule has been observed. The recovery of Rg was a few percent. An  enrichment factor of three to four orders of magnitude has been obtained. It is deduced that under our experimental conditions the evaporation rate of Rg is lower than that of Au.

16 mg of 99.999\% Au metal, supplied by Leico Industries,   were held in a quartz tube in vacuum for two weeks at a temperature of 1127$^\circ$ C (63$^\circ$ C above the melting point of Au). The heating was stopped when the residue could no longer be seen with a magnifying glass. The minute residue was dissolved overnight in aqua regia at 60$^\circ$ C. After evaporating almost to dryness, 3 M HNO$_{3}$ was added and diluted to obtain a final volume of 20 ml at  0.7 M HNO$_{3}$.  This solution  was   studied with the ICP-SFMS.

                    \begin{figure}[h]
\vspace*{-0.2 cm}
\hspace*{-0.7cm}

\includegraphics[width=0.48\textwidth]{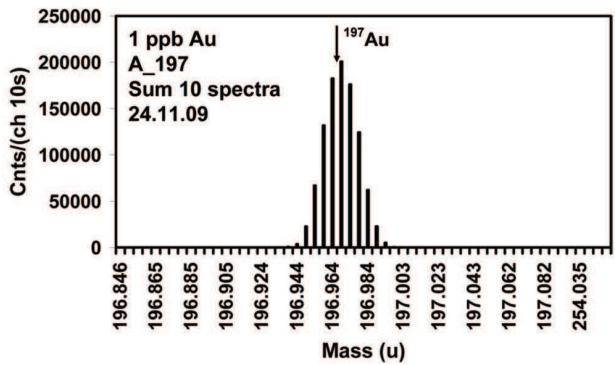}
\vspace*{-0.6 cm}
 \caption{Measurements of mass  197
 obtained with 1 ppb Au solution.
 The sum of ten measurements is displayed. Total measuring time: 700
 s. Observed mass
 position: 196.969 u. Known mass of $^{197}$Au (indicated by an arrow)  is 196.967 u \cite{aud03}.}
\end{figure}

           \begin{figure}[h]
\vspace*{-0.6 cm}
\includegraphics[width=0.48\textwidth]{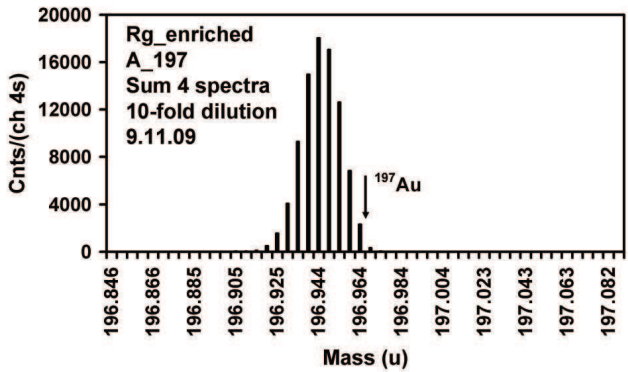}
\vspace*{-0.6 cm}
 \caption{Measurements of mass  197
 obtained with a ten-fold diluted sample of the enriched Rg  solution.
 The sum of four measurements is displayed. Total measuring time: 280
 s. Observed mass
 position: 196.945
  u. Known mass of $^{197}$Au (indicated by an arrow)  is 196.967 u \cite{aud03}.}
\end{figure}

The accurate mass measurements were similar to those described before  \cite{mar07,mar09,mar08}. The ICP-SFMS was a Finnigan Element2 (Thermo-Electron, Bremen, Germany). In this instrument a solution
of the material to be studied is introduced into a high
temperature (6000 - 8000 K) plasma source. At these temperatures
predominantly atomic species are present. Molecular ions are
formed after the source, mainly by interaction with oxygen and
hydrogen ions from the solution. The predefined  resolution
mode, m/$\Delta$m = 4000 (10\% valley definition), was used
throughout the experiments to separate atomic ions from molecules
 with the same mass number. The sensitivity-enhanced setup of the
instrument was similar to that described in \cite{rod04}.
In the present experiment the sample uptake rate was 50 $\mu$l
  min$^{-1}$ and the
 sensitivity for $^{197}$Au in this resolution mode was 2x10$^{7}$ counts
  s$^{-1}$mg$^{-1}$l$^{-1}$. Methane gas was added to the plasma to decrease the formation of
  molecular ions \cite{rod05}. Oxide and hydride formation (monitored
  as UO$^{+}$/U$^{+}$
  and UH$^{+}$/U$^{+}$ intensity ratios) were approximately 0.04 and 1x10$^{-5}$,
   respectively.
   Mass calibration was performed  using the $^{115}$In$^{+}$,
   $^{232}$Th$^{+}$, $^{235}$U$^{+}$,
   $^{238}$U$^{+}$ and $^{238}$U$^{16}$O$^{+}$ peaks.

    Complete elemental screening was performed on the enriched Rg solution to
       assess the impurity levels. The amount of
       certain trace elements with relatively high concentrations, expressed as ppb (g/g)  of the solution are as follows:

       Au 4.1, Ca 140, Cl 15600, K 250, Na 360, P 150, S 780, Si 40, Th 0.001, U 0.04.

                  \begin{figure}[h]
\includegraphics[width=0.48\textwidth]{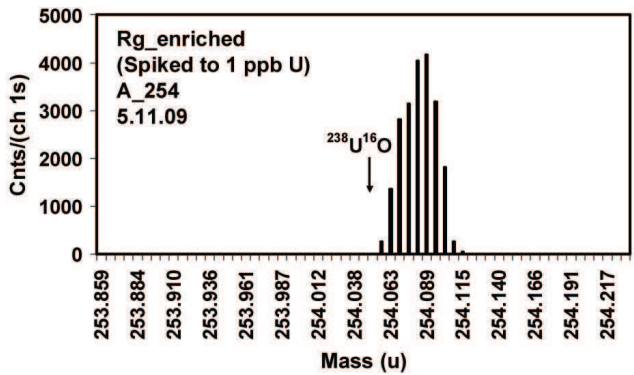}
\vspace*{-0.5 cm}
\caption{Measurement of mass  254
 obtained with   the enriched Rg solution which was spiked to 1 ppb U (see text). Total measuring time: 70
 s. Observed mass
 position: 254.084 u. It is shifted from the known mass
   of $^{238}$U$^{16}$O of 254.046 u \cite{aud03} (indicated by an arrow) by 0.038 u.}

 \end{figure}

        These concentrations are correct to within a factor of two. After these impurity determinations, the solution was spiked to 1 ppb of U. These levels of contamination are higher than those specified in  \cite{mar09} where pure original Au solution, without processing, was used. (The corresponding values there for Na and S were in the region of a few ppb.) A synthetic solution  which contains  the above-mentioned impurities with their measured concentrations, spiked to 1 ppb of U, was prepared and used for comparison in the experiments.

        After a two-hour warm-up, the machine was still unstable. Therefore the measurements were started after nine hours of stabilization.    A range of 0.45 u, divided into 70 channels, was scanned in each
  spectrum. The scanning was performed by changing the acceleration voltage while keeping the magnetic field constant.
 The  mass 261 was analyzed
 with an integration time per channel of 1 s and run 90 times. The calibration was checked by measuring mass 254 ($^{238}$U$^{16}$O) before and after the long series of  measurements on mass 261.   A calibration correction of 0.037 u was deduced from these measurements (see below) and applied to the data.

Figure 2 shows the sum spectrum of ten measurements of mass
 197 ($^{197}$Au) taken with a 1 ppb (g/g) Au solution.  The FWHM of the peak is 0.038 u.

 Figure 3 shows the sum of four measurements of mass
 197, taken with a ten-fold diluted  sample of the enriched Rg solution.  The peak of  $^{197}$Au is
shifted from its known value \cite{aud03} by $-$0.022 u. By a comparison of Figs. 2 and 3 it is concluded that the Au concentration in the enriched
  Rg solution is 2.2 ppb (g/g). It is therefore deduced that the residue of Au in this 20 ml solution is 44 ng which is about 3x10$^{-6}$ of its initial value of 16 mg.

Figure 4 shows the results obtained on mass 254 measured with the enriched Rg solution which was spiked to 1 ppb U. It was taken before performing the  measurements on mass 261.   The peak of $^{238}$U$^{16}$O is seen and it is shifted from the known  value \cite{aud03}  by 0.038 u. A similar spectrum taken after  the  mass region 261 measurements were finished gave a shift of 0.036 u. An average correction of 0.037 u was applied to the data.

Figure 5 shows the sum of 90 spectra obtained on mass 261 with the enriched Rg solution (a) and with the synthetic solution (b). Figure 5(c) presents the subtraction of spectrum (b) from spectrum (a).
A peak of 37$\pm$13 counts is seen in Fig. 5(c) at the predicted position of $^{261}$Rg \cite{mol95,lir00,kou05}. The error estimate was calculated according to the formula $\sigma$=(N$_{total}$+N$_{back}$)$^{1/2}$ \cite{bev69} where N$_{total}$ is the total number of counts in the Rg region taken from Fig. 5(a) and N$_{back}$ is the number of background counts taken from Fig. 5(b).

          \begin{figure}[h,t]
\includegraphics[width=0.48\textwidth]{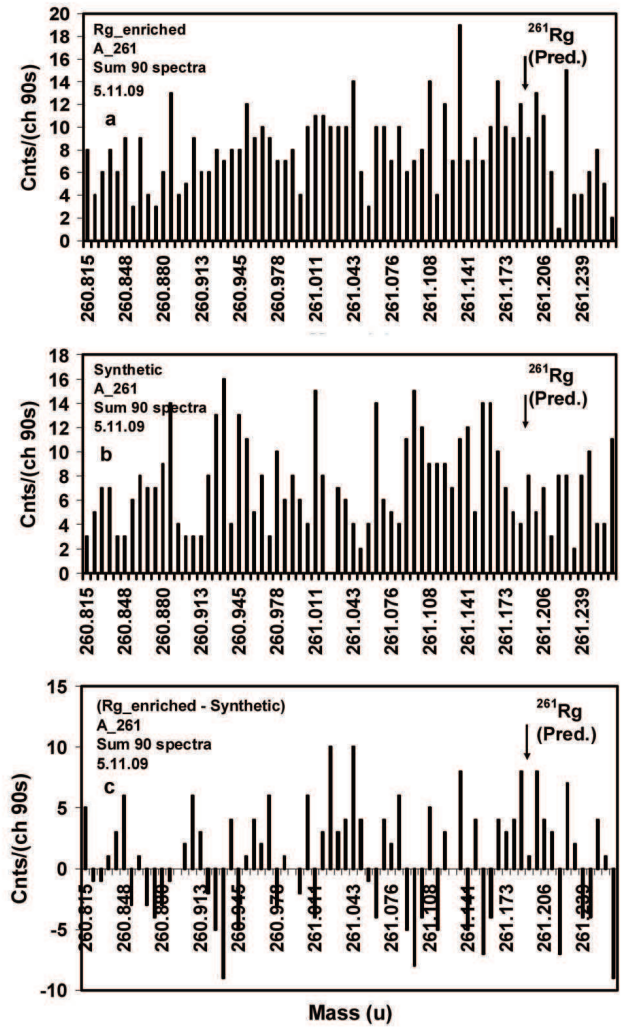}
\vspace*{-0.8 cm}
\caption{Measurements of mass  261
 obtained with the enriched Rg solution (a), and with the synthetic solution (b). Spectrum (c) represents the subtraction of spectrum (b) from spectrum (a).
 The sum of 90 spectra is displayed in each case. Total measuring time: 6300
 s in spectra (a) and (b). The arrows indicate the predicted position of $^{261}$Rg (m=261.154 u)   \cite{mol95,lir00,kou05} shifted by a calibration correction of 0.037 u (see text).}

 \end{figure}

The 90 measured  spectra are independent of one another. Each   measurement used a different sample of the solution. At low statistics at an abundance level of about 10$^{-15}$ of the solution (see below) it is possible that in some spectra the peak of $^{261}$Rg will be more pronounced than in others.

\begin{widetext}

            \begin{figure}[h]

\vspace*{-0.4 cm}
\includegraphics[width=0.85\textwidth]{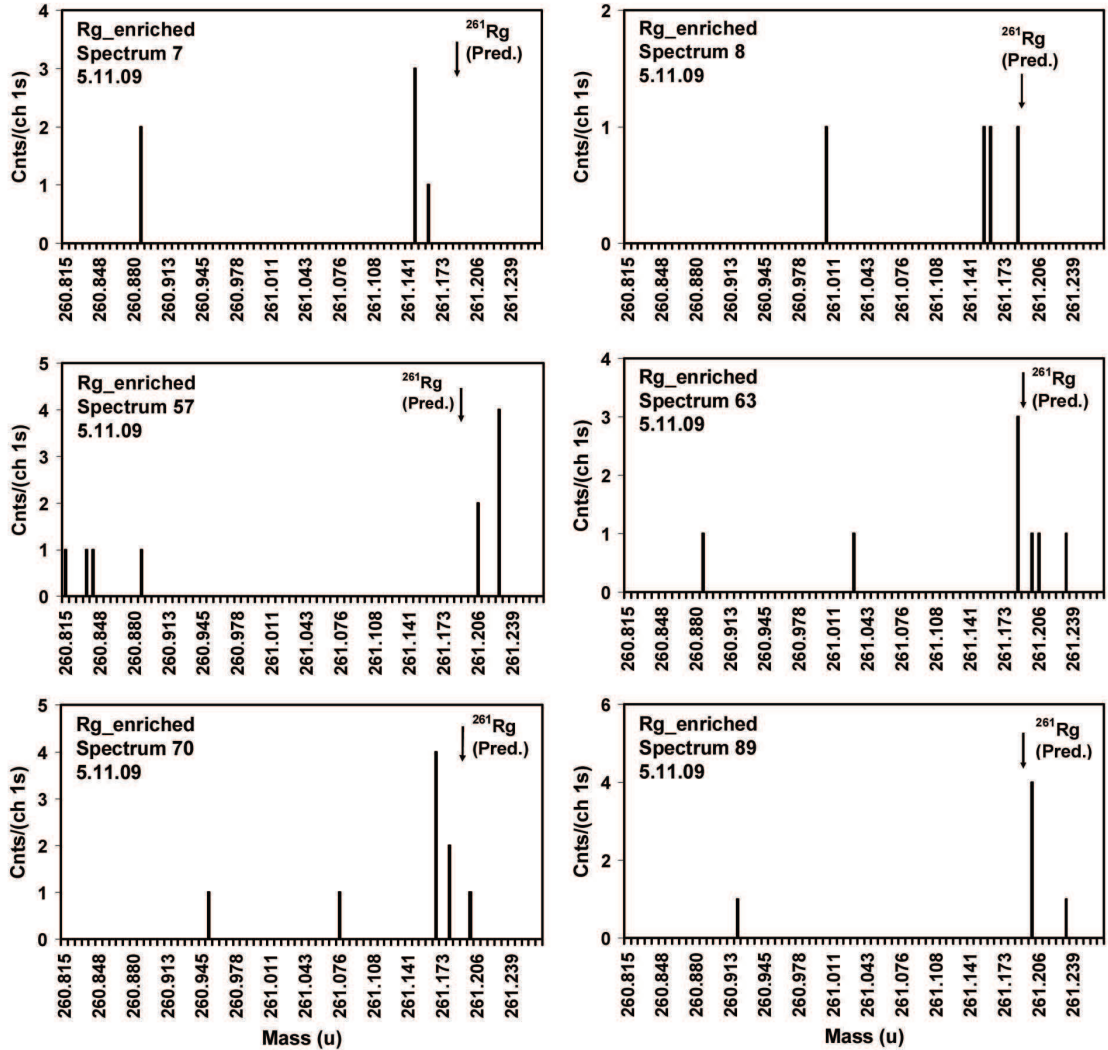}
\vspace*{-0.4 cm}
\caption{Six individual spectra at the mass region  261
 obtained with the enriched Rg solution are shown.  Total measuring time of each spectrum is 70 s. The arrows indicate the predicted position of $^{261}$Rg (m = 261.154 u)   \cite{mol95,lir00,kou05} shifted by a calibration correction of 0.037 u (see text).}

 \end{figure}
\vspace*{-0.4cm}

\end{widetext}

\vspace*{-0.7cm}

 Such  behavior has been seen by us for instance in a study of the known molecule $^{232}$Th$^{40}$Ar$^{16}$O at 0.1 ppm of Th, where in one spectrum 12 counts were seen  and in the spectrum measured immediately afterward,  one count was observed. In the spectrum of Fig. 5(a) 130 counts are seen in the region of interest (ROI) of $\pm$0.044 u around the predicted position of $^{261}$Rg. In the region out of interest (OROI) from 260.815 to 261.147 u (3.8 times larger than the ROI) there are  420 counts. If the group of events around the predicted position of $^{261}$Rg is due to background, then the ratio N$_{OROI}$/N$_{ROI}$ = 3.23 should be the same, within statistics, in each individual spectrum. In 26 spectra events were seen at the predicted mass of $^{261}$Rg. In 11 spectra it was found that this  ratio is statistically  different from 3.23, between 0.2 and 1.4 with an average value of 0.6.  These spectra are dominated by counts around the predicted position of $^{261}$Rg. The data  are as follows: (Spectrum number; N$_{ROI}$; N$_{OROI}$; N$_{OROI}$/N$_{ROI}$; P): (1; 2; 1; 0.5; 1x10$^{-2}$), (7; 4; 2; 0.5; 2x10$^{-4}$ ), (8; 3; 1; 0.3; 6x10$^{-4}$), (27; 2; 1; 0.5; 1x10$^{-2}$), (57; 6; 4; 0.7; 2x10$^{-5}$), (63; 6; 2; 0.3; 7x10$^{-7}$), (69; 4; 3; 0.8; 9x10$^{-4}$), (70; 7; 2; 0.3; 4x10$^{-8}$), (79; 5; 7; 1.4; 5x10$^{-3}$), (84; 4; 4; 1.0; 3x10$^{-3}$), (89; 5; 1; 0.2; 2x10$^{-6}$). P is the Poisson probability of observing N$_{OROI}$ events when N$_{ROI}$ x 3.23 are expected. As seen these probabilities are between 1x10$^{-2}$ to 4x10$^{-8}$ for each of these  individual spectra.   These results are  displayed  in Figs. 6 and 7.

In Fig. 6, six out of the 11  mentioned  spectra are seen. In these as well as in most of the 11 spectra, more events are seen in the  high mass region, around the predicted  $^{261}$Rg position, than in the rest of the measured region.

                    \begin{figure}[t]

\vspace*{-0.1 cm}
\hspace*{-0.7cm}

\includegraphics[width=0.48\textwidth]{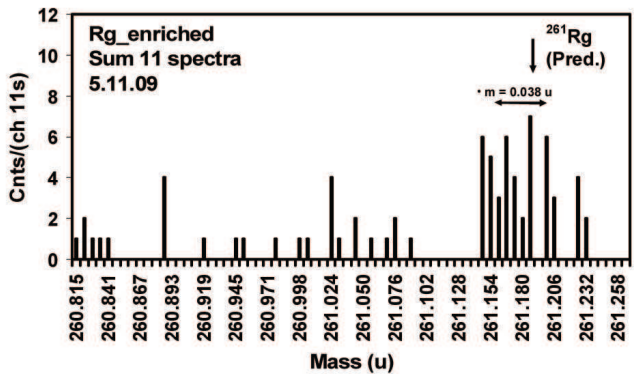}
\vspace*{-0.6 cm}
 \caption{Measurements of mass  261
 obtained with the enriched Rg solution.
 The sum of 11 measurements is displayed. Total measuring time: 770
 s. Observed mass
 position, corrected by a calibration shift of 0.037 u, is 261.144$\pm$0.020 u. Predicted mass \cite{mol95,lir00,kou05} for $^{261}$Rg (indicated by an arrow)  is 261.154 u.}
\end{figure}

In Fig. 7 the sum of the 11  spectra is shown. A pronounced peak of 48 counts is seen at mass 261.144$\pm$0.020 u (taking into account the calibration shift).  The total number of events OROI is 28 as compared to 155 (48 x 3.33) deduced if the ratio of N$_{OROI}$/N$_{ROI}$ = 3.23 of Fig. 5(a) is respected. (The probability of seeing  28 counts when 155 are expected is extremely small.) The observed mass of this peak fits, within 0.010 u, with the predicted mass \cite{mol95,lir00,kou05} of $^{261}$Rg of 161.154 u. It is larger than the mass of any molecule with  M = 261 (except for multi-H, -Li, -Be and -B molecules) as seen in Fig. 1.  By a comparison with Fig. 3  the average  intensity of this peak in the 90 spectra, compared to $^{197}$Au is about 2x10$^{-6}$ (4x10$^{-15}$ of the solution). It is about a factor of ten larger in the 11 spectra mentioned above, about 4x10$^{-14}$ of the solution.  Thus, an enrichment of three to four orders of magnitude, compared to the earlier determined abundance of (1-10)x10$^{-10}$ \cite{mar09}, has been achieved  in this experiment. The recovery of Rg from the initial Au is estimated to be a few percent. The high enrichment factor shows that the evaporation rate  of Rg in a matrix of Au is lower than that of Au.

In summary, the present results strengthen the first
observation \cite{mar09} of the superheavy element Rg (Z = 111)
in nature. As was argued in \cite{mar09}, if its terrestrial concentration about 4.5x10$^{9}$ y ago
was similar to that of Au,
then the half-life of the observed $^{261}$Rg nucleus is $\geq$ 10$^{8}$ y.
Since the half-life of the normally deformed g.s. of $^{261}$Rg \cite{mol97}
 is predicted to be about 1 $\mu$s, it was concluded \cite{mar09}
that the observed nucleus is in an isomeric state, probably unrelated to high-spin states near closed shells, ¯fission
isomers or K-isomers at normal deformations. It is therefore
hypothesized \cite{mar09} that it belongs to a new class
of high spin isomeric states in the SD or HD minimum
of the nuclear potential energy \cite{mar96a,mar96b,mar01a,mar01b}. This work also
shows that it is possible to enrich Rg in purified Au by
evaporating the Au in a vacuum at about 63$^{o}$ C above
its melting point.

We appreciate valuable discussions and help of N. Zeldes, E. Grushka, U. Kaldor and J.L. Weil. We thank I. Rodushkin for performing  the ICP-SFMS measurements and  I. Felner and M. Tsindlekht for helping with  evaporating the Au.

\end{document}